\documentclass[]{spie}

\usepackage[]{graphicx}
\usepackage{amsmath}
\usepackage{amssymb}
\usepackage[]{aas_macros}
\usepackage{bm}
\usepackage{ulem}
\usepackage{cancel}

\def\ra{\mathrm{a}}

\def\rb{\mathrm{b}}

\def\rc{\mathrm{c}}
\def\bC{{\bf C}}
\newcommand{\rd}{\mathrm{d}} 
\newcommand{\rD}{\mathrm{D}}

\def\rg{\mathrm{g}}
\def\bG{{\bf G}}
\def\rG{\mathrm{G}}

\def\bH{{\bf H}}
\def\rH{\mathrm{H}}

\def\rk{\mathrm{k}}

\def\rm{\mathrm{m}}
\def\bm{{\bf m}}
\def\bM{{\bf M}}

\def\rn{\mathrm{n}}

\def\rp{\mathrm{p}}

\def\bQ{{\bf Q}}
\def\br{\boldsymbol{r}}
\def\rr{\mathrm{r}}

\def\rT{\mathrm{T}}
\def\ru{\mathrm{u}}

\def\bv{{\bf v}}
\def\bV{{\bf V}}
\def\bw{{\bf w}}
\def\rw{\mathrm{w}}

\def\bx{{\bf x}}
\def\by{{\bf y}}
\def\phihf{\phi_\mathrm{hf}}
\def\balpha{\boldsymbol{\alpha}}
\def\bchi{\boldsymbol{\chi}}
\def\bPsi{\boldsymbol{\Psi}}

\def\brho{\boldsymbol{\rho}}

\def\boldeta{\boldsymbol{\eta}}

\def\bnu{\boldsymbol{\nu}}

\title{A Statistical Framework for Utilization of Simultaneous Pupil
  Plane and Focal Plane Telemetry for Exoplanet Imaging, Part II: The
  Science Camera Image as a Function of the Wavefront Sensor Field}

\author{Richard A. Frazin
\skiplinehalf
{\small Dept. of Climate and Space Sciences and Engineering, University of Michigan, Ann Arbor, MI 48109} }
\authorinfo{E-mail: rfrazin {\it \_at\_} umich.edu}

\begin{document}
\maketitle

\begin{abstract}

In an effort to transcend the limitations of differential imaging of exoplanets in the era of extremely large telescopes (ELTs),
the first paper in this series established a rigorous, fully polarimetric framework for determining the science camera (SC) image given a turbulent wavefront and unknown aberrations in multiple planes the optical system. 
This article builds on the structure developed in Paper I in order to rigorously express the polarimetric SC image in terms of the field impinging on the wavefront sensor (WFS), thereby providing a direct connection between the measurements made in both subsystems.
This formulation allows the SC image to be written as a function of the WFS measurements, including the following unknown quantities which can, in principle, be estimated via statistical inference: the non-common path aberration (NCPA), WFS gain errors, aberrations downstream of the beamsplitter, and the planetary image.
It is demonstrated that WFS bias error is mathematically equivalent to NCPA.
Thus, with the ability to treat WFS bias and gains, the method should not be overly sensitive to WFS calibration problems.
Importantly, this formulation includes stochastic processes that represent noisy measurement of the SC image, noisy WFS measurements, and high-frequency components of the wavefront to which the WFS is insensitive.
It is shown that wavefront error due to noise in the WFS measurements has a convenient semi-analytical representation in terms of the WFS measurement operator's singular functions.   
Further, the first and second order statistics of these processes are specified, thereby setting the stage for the application of statistical inference methods to be describe in later papers in this series.  

\end{abstract}

\section{Introduction}

This is the second installment of a series of articles dedicated to development of rigorous estimation methods for the ultra-high contrast astronomical imaging problem for ground-based observatories, such as is encountered when attempting to make direct images of exoplanetary systems.
While the discussion has some relevance to space-based platforms, the atmospheric modulation of the wavefronts and the partial demodulation by adaptive optics (AO) systems present an essential difference.
The flat wavefront encountered in space provides the possibility for highly effective coronagraphy, so that the starlight is almost entirely extinguished by the hardware, and it remains to differentiate between the effects of unknown optical aberrations and the planetary emission \cite{Traub_Review}.
Determining the unknown optical aberrations, or at least somehow treating their manifestation as speckles in the image plane, has been the subject of extensive literature.
In particular, space-based imaging schemes are going to greatly benefit from "dark hole" or electric field conjugation (EFC) techniques, which utilize focal plane sensing procedures and a deformable mirror (DM) to further extinguish the starlight in a specified region of the image [\citenum{Traub_Nulling06,Pueyo_EFC09}].
However, ground based observations, even those employing so-called "extreme" (i.e., high-order) AO, cannot reject the starlight nearly as efficiently because coronagraphs only work optimally when presented with a flat wavefront. 
Further, in the first paper in this series (henceforth Paper I) [\citenum{Frazin16a}], the author demonstrated that that the "dark hole" and EFC methods are likely to be far less effective than in space due to the rapidly changing nature of the turbulent wavefront, as these methods can only treat aberrations that are in planes optically conjugate to the DM when confronted with a highly time-dependent wavefront.
This due to fact that they implicitly rely on the assumption of equivalent aberrations, which states that all aberrations in the optical system can be considered equivalent to aberrations in a single plane (a time-independent wavefront, as encountered in space, makes this assumption moot).

While a turbulent wavefront presents a variety of difficulties, it also provides a valuable opportunity to interrogate the optical system by using the wavefront sensor (WFS) measurements to make an explicit determination of the optical aberrations that give rise to the speckle field.  
In previous works, the author demonstrated that is is possible to use simultaneous millisecond exposures by the science camera (SC) and the WFS to determine both the planetary image and the non-common path (pupil plane) aberrations (NCPA) \cite{Frazin13,Frazin14}.
Importantly, this approach is fully compatible with other sources of information such as constraints from diurnal rotation and multi-wavelength observations, such as are exploited by differential imaging techniques, including polarization differential imaging \cite{Hinkley_PDI09}.
The current methodologies based on differential imaging and their attendant systematic errors are reviewed in Paper I and [\citenum{Frazin13,Rameau_ADI_SDI_limits15}].

In an effort to transcend the limitations of differential imaging in the era of extremely large telescopes (ELTs),
Paper I established a rigorous, fully polarimetric framework for determining the SC image given a turbulent wavefront and unknown aberrations in multiple planes the optical system.   
The polarimetric treatment is necessary due to the polarizing effects of the optical system [\citenum{Breckinridge15}], which promises to undermine the effectiveness of differential imaging schemes for the most ambitious targets.
This article builds on the structure {\it mise en place} in Paper I in order to rigorously express the SC polarimetric image in terms of the field impinging on the WFS, thereby providing a direct connection between the measurements made in both subsystems.
This formulation allows the SC image to be written as a function of the WFS measurements, including the following unknown quantities which can, in principle, be estimated via statistical inference: the NCPA, WFS gain errors, aberrations downstream of the beamsplitter, and the planetary image.
It is demonstrated that WFS bias error is mathematically equivalent to NCPA.
Thus, with the ability to treat WFS bias and gains, the method should not be overly sensitive to WFS calibration problems.
Importantly, this formulation includes stochastic processes that represent noisy measurement of the SC image, noisy WFS measurements, and high-frequency components of the wavefront to which the WFS is insensitive. 
It is shown that wavefront error due to noise in the WFS measurements has a convenient semi-analytical representation in terms of the WFS measurement operator's singular functions. 
Further, the first and second order statistics of these processes are specified (or at least referred to the literature), thereby setting the stage for the application of statistical inference methods to be described in later papers in this series.  

While the author has attempted to make this discussion self-contained, the reader is expected to have familiarity with material presented in Paper I.

\section{Propagation Within the Optical System}\label{OpticalSystem}

Consider a telescope with a closed-loop AO system, as shown in Fig.~3 of Paper I.   
As per the discussion in Paper I, the star at the center of the putative planetary system gives rise to an electromagnetic disturbance impinging on the telescope entrance pupil that can be represented as:
\begin{equation}
u_0(\br_0,t) = \sqrt{I_\star \,} \exp j \big[  k \balpha_\star \cdot \br_0 + \phi_\ra(\br_0,t) \big] 
\breve{u}_\star \, ,
\label{u_0}
\end{equation}
where $\br_0$\ is the coordinate in the telescope entrance pupil plane, $\balpha_\star$\ is the (small) sky-angle of the star relative to the telescope pointing direction, $k= 2 \pi / \lambda $\ is the scalar-valued wavenumber corresponding to wavelength $\lambda$, $\phi_\ra(\br_0,t)$\ is the complex-valued atmospheric modulation, and $\breve{u}$\ is the polarization state vector of the starlight impinging on the atmosphere (see Sec. 2.A of Paper I).

\subsection{DM and Residual Phase}\label{DM}

Using operator notation, the stellar field incident on the DM surface, denoted with index $D$, is given by:
\begin{equation}
u_D(\br_D,t) \equiv u_D^-(\br_D,t) 
 = \sqrt{I_\star} \Upsilon_{D,0}(\br_D,\br_0)\breve{u}_\star
\exp j \big[ k \balpha_\star \cdot \br_0 + \phi_\ra(\br_0,t)  \big]  \, ,
\label{u_D-} 
\end{equation}
where the $^-$\ superscript emphasizes that field has not yet interacted with the DM surface, $\Upsilon_{D,0}$\ is a $2 \times 2$\ matrix-valued propagation operator that relates the field at the telescope entrance (plane $0$) to the field at the DM (plane $D$), and $\br_D$\ is the coordinate on the DM surface.
Note that the equivalence of $u_D(\br_D,t)$\ and $u_D^-(\br_D,t)$\ is in keeping with notational conventions established in Part I.

The action of the DM can be described by a Jones pupil matrix $A_D\big( \br,\bm(t) \big)\exp [-j\mu\big(\br , \bm(t) \big) ]$, in which $\bm(t)$\ is the vector of mirror command positions at time $t$, $A_D\big( \br , \bm(t) \big)$\ is a matrix accounting for the polarization aberration imparted by the DM and $ \exp [-j\mu\big(\br,\bm(t) \big) ]$\ accounts for the optical path difference that the DM applies.
The function $- \mu \big( \br , \bm(t) \big)$\ is the phase perturbation that the DM imparts to the wavefront and it is given by $\mu \big(\br , \bm \big) =  4 \pi h \big(\br , \bm \big)/\lambda$, where $h(\br , \bm)$\ is the height function, which is often approximated as being linear in $\bm$. 
The minus sign is chosen because, as per Part I, the light is propagating in the positive (local) $z$ direction and a positive mirror displacement causes a retardation in the phase of the reflected wave.
As the DM displacements are small and the reflective surface is smooth, the effect of the DM displacements on the polarization of the wave is likely negligible, so the $\bm(t)$\ argument can be dropped from $A_D$, i.e., $A_D \big( \br_D,\bm(t) \big) \approx A_D(\br_D)$.   
The state of the field immediately after reflection off the DM is given by:
\begin{equation}
u_D^+(\br_D,t) = A_D ( \br_D ) u_D(\br_D,t)  \exp [-j\mu\big(\br_D,\bm(t) \big) ] \, ,
\label{DM_action}
\end{equation}
where the $^+$\ superscript indicates that the field has reflected off of the DM.
Using Eqs.~(\ref{u_D-}) and (\ref{DM_action}), one has:
\begin{equation}
u^+_D  (\br_D, t)   =  \sqrt{I_\star \,} A_D ( \br_D ) \Upsilon_{D,0}(\br_D,\br_0) \breve{u}_\star
 \exp  j \big[ k \alpha_\star \cdot \br_0 +   \phi_\ra(\br_0,t) - \mu\big(\br_D,\bm(t) \big)  \big] \, .
\label{u_D+closed}
\end{equation}
The following overloaded definition of the propagation operator $\Upsilon_{D+1,D}$\ (where the $D+1$\ surface may be the beam splitter) is convenient:
\begin{equation}
\Upsilon_{D+1,D}(\br_{D+1},\br_D;\bm(t)) = 
\Upsilon_{D+1,D}(\br_{D+1},\br_D)\exp \big[ -j \mu \big(\br_D,\bm(t) \big)  \big] \, . 
\label{Upsilon-DD+1}
\end{equation}
Thus, the propagation operators $\Upsilon_{D+1,D}(\br_{D+1},\br_D;\bm(t))$ and $\Upsilon_{D+1,D}(\br_{D+1},\br_D)$\ include the Jones pupil matrix $A_D(\br_D)$\ accounting for the polarization effects of the mirror at zero DM displacement.
As per the usual definition of the propagation operators, this operator relates the field incident on the DM surface (index $D$) to the field incident on the next optical surface, indexed by $D+1$.

In common parlance, the residual phase is the part of the atmospheric modulation that the DM does not correct.
However, the rigors of ultra-high contrast imaging require us to examine the light impinging on the DM in more detail.  
In Eq.~(\ref{u_D+closed}), the reader should take particular note of the fact that $\phi_\ra(\br_0,t)$\ and $\mu \big( \br_D, \bm(t) \big)$\ are not functions of the same spatial argument; $\br_0$ refers to the coordinate in the telescope entrance pupil plane, and $\br_D$\ refers to the DM plane.
Indeed, little can be said about the effect of the DM on the beam without specifying the nature of the $\Upsilon_{D,0}$\ operator, which propagates the field from the telescope entrance to the DM surface.

The $0$\ plane is a pupil plane, and in order to correct the turbulent modulation of the wavefront, the $D$\ plane must be conjugate to it, making it a pupil plane as well.
(In multi-conjugate AO systems, additional DMs are conjugate to various planes that are a number of kilometers above the telescope.) 
Ideally, that is, assuming this conjugate relationship, and ignoring Fresnel propagation effects, obscurations (e.g., spiders and masks) and aberrations, Eq.~(\ref{u_D-}) specializes to a demagnified version of Eq.~(\ref{u_0}):
\begin{equation}
u_D^\mathrm{ideal}(\pm \, \br,t)  = \beta u_0(\beta \br,t) 
= \beta \sqrt{I_\star }  \exp j \big[ \beta k \balpha_\star \cdot \br  
+ \phi_\ra(\beta \br,t) \big] \breve{u}_\star  \, ,
\label{demagnify_ideal}
\end{equation}
 in which $\beta$\ is the ratio of the telescope entrance aperture diameter to the diameter of the beam impinging upon the WFS.  
For example, the Subaru telescope has a primary mirror diameter of $8.2$\ m and the diameter of the beam hitting the DM in SCExAO coronagraphic imaging system is $18$ mm, leading to a value of $\beta \approx 455$ [\citenum{Guyon_SCExAO13}].
The orientation, represented by the sign in $\pm \br$, depends on how many times the pupil plane has been conjugated. 
Henceforth, a positive orientation, corresponding to an even number of conjugations of the telescope entrance pupil, will be assumed to avoid cumbersome notation.
Using Eq.~(\ref{demagnify_ideal}) on the right hand side of Eq.~(\ref{DM_action}), it is readily seen that the residual phase $\phi_\rr(\br,t)$\ can be defined as $\phi_\rr^{\mathrm{ideal}}(\br,t) =  \phi_\ra(\beta \br,t) - \mu \big( \br, \bm(t)   \big)$, in keeping with the standard, if simplistic, usage of the term.

In order to understand the meaning of the residual phase in a real optical system, instead of Eq.~(\ref{demagnify_ideal}), one may write: 
\begin{equation}
u_D  (\br, t)  = \beta \sqrt{I_\star \,}
   T_{D,0} \big( \br, \phi_\ra(\beta \br,t)  \big) \breve{u}_\star
 \exp  j \big[ \beta k \balpha_\star \cdot \br +  \phi_T \big( \br,  \phi_\ra(\beta \br,t) \big) + \phi_\ra(\beta \br,t)  \big] \, ,
\label{u_D_arb}
\end{equation}
where the Jones pupil matrix  $T_{D,0} \big( \br, \phi_\ra(\beta \br,t)  \big)$ accounts for the polarization effects of the telescope system up to the $D$\ plane, and $\phi_T \big( \br, \phi_\ra(\beta \br,t)  \big)$\ accounts for the effects of scalar aberrations between the $0$ and $D$ planes.
It is important to emphasize that the scalar aberration $\phi_T$\ must have a dependence on the atmospheric modulation $\phi_\ra$, in keeping with the invalidity of equivalent aberrations assumption, as discussed in Sec.~4A.1 of Paper I.   
As a number of the aberrations likely occur on surfaces that are not conjugate to the DM plane, there is no time-independent aberration in the DM plane that is their equivalent, as was discussed in Sec. 4.A.1 of Paper I.
Similar comments apply to the function  $T_{D,0} \big( \br, \phi_\ra(\beta \br,t)  \big)$.
The arbitrary nature of $T_{D,0}$\ and $\phi_T$\ functions make Eq.~(\ref{u_D_arb}) fully general, but retaining the some of functional form of Eq.~(\ref{demagnify_ideal}) is useful for pedagogical purposes.
The residual phase, $\phi_\rr$, can be defined as:
\begin{equation}
\phi_\rr(\br,t)   \equiv    \phi_\ra(\beta \br,t) + 
\phi_T \big( \br,  \phi_\ra(\beta \br,t) \big) - \mu\big(\br,\bm(t) \big) \, .
\label{residual_phase_def}
\end{equation}
With current DM technology, $\mu \big(\br,\bm(t) \big)$\ in Eq.~(\ref{residual_phase_def}) is constrained to be a real function, so the DM control loop attempts to cancel the real part of $\phi_\ra + \phi_T$, resulting in a flatter wavefront.   
It is expected that, for most locations $\br$, $ | \phi_T \big( \br, \phi_\ra(\br,t) \big) | < | \phi_\ra(\br,t) |$ since $\phi_T$\ is due to presumably small aberrations. 
Unless the telescope exhibits rapid ($\sim$kHz) vibrations, the temporal bandwidth of $ \phi_T \big( \br, \phi_\ra(\br,t) \big)$\ should be same as that of $ \phi_\ra(\br,t)$, however, $\phi_T$\ may exhibit substantially more high spatial-frequency content than $\phi_\ra$, depending on the character of the aberrations.
As the functions $T_{D,0} \big( \br, \phi_\ra(\beta \br,t)  \big)$\ and $\phi_T \big( \br,  \phi_\ra(\beta \br,t) \big)$\ are difficult to to determine, the developments to follow will not utilize them.

\subsection{Wavefront Sensor and Science Camera}\label{WFSandSC}

In order to utilize the pupil and focal plane telemetry, it is important to understand the relationship between the fields impinging on the SC and the WFS.
As can be seen in Fig.~3 of Paper I, after interacting with the DM the light goes through a beam splitter (BS), with some going to the WFS and some to the coronagraph and SC.  
Let the BS, WFS, SC detector planes be denoted with the indices $B$, $w$ and $C$, respectively.
Then, the WFS and SC fields arising from the star can be expressed as:
\begin{align}
u_w(\br_w,t) &= \Upsilon_{w,B}(\br_w,\br_B) \Upsilon_{B,D}(\br_B,\br_D) 
 \Upsilon_{D,0}(\br_D,\br_0)  
  \breve{u}_\star 
\exp  j \big[ k \alpha_\star \cdot \br_0 +   \phi_\ra(\br_0,t) - \mu\big(\br_D,\bm(t) \big)  \big]  
\label{u_w_complete}\\ 
 u_C(\br_C,t) & = \Upsilon_{C,B}(\br_C,\br_B) \Upsilon_{B,D}(\br_B,\br_D) 
  \Upsilon_{D,0}(\br_D,\br_0)   \breve{u}_\star
\exp  j \big[ k \alpha_\star \cdot \br_0 +   \phi_\ra(\br_0,t) - \mu\big(\br_D,\bm(t) \big)  \big] \, .
\label{u_C_complete}
\end{align}
Both Eqs.~(\ref{u_w_complete}) and (\ref{u_C_complete}) implicitly make use of Eq.~(\ref{Upsilon-DD+1}).
Note that the only difference in these two expressions is the leftmost operator, which propagates the light from the BS to the WFS in Eq.~(\ref{u_w_complete}), and from the BS to the SC in Eq.~(\ref{u_C_complete}).
As will be discussed shortly, it is likely that the operator $ \Upsilon_{w,B}$\ is invertible (or nearly so), allowing one to write $u_C$\ in terms of $u_w$, which is useful because the latter is measured by the WFS.
The operator $\Upsilon^{-1}_{w,B}(\br_B,\br_w)$ (noting the order of the arguments) is back-propagation operator that calculates the field in plane in the BS plane ($B$) from the field in the WFS plane ($w$).
Placing $\Upsilon^{-1}_{w,B}(\br_B,\br_w)$\ on the left of both sides of Eq.~(\ref{u_w_complete}), it is easy to see that Eq.~(\ref{u_C_complete}) can be rewritten as:
\begin{equation}
 u_C(\br_C,t) = \Upsilon_{C,B}(\br_C,\br_B) \Upsilon^{-1}_{w,B}(\br_B,\br_w) u_w(\br_w,t)\, .
\label{u_C-u_w-2}
\end{equation}
The relation shown in Eq.~(\ref{u_C-u_w-2}) depends on the invertibility of the $\Upsilon_{w,B}$ operator.
If it is nearly invertible, then an approximate inverse can be used, and the consequences of the approximation must be assessed.  
The $B$ and $w$ planes are both pupil planes, and therefore are conjugate.  Further, they are likely to be located near each other, reducing the importance of Fresnel propagation.  

Assuming no further magnification, under geometrical optics, a reasonable model for $\Upsilon_{w,B}$\ is:
\begin{equation}
\Upsilon_{w,B}(\br_w,\br_B) = \delta\big( \br_w - \br_B \big) \big[ \mathbb{I} + A_w^\rk(\br_w)  \big] 
\exp [ - j \phi_\ru(\br_w)  ] \, , 
\label{Upsilon_wB}
\end{equation}
in which $\phi_\ru$\ accounts for unknown scalar aberrations between the BS and the WFS.  
Since the beam between these two planes is expected to be columnated, representing the aberrations on the intervening optical surfaces as a single one should be adequate, as per Sec. 4.A.1 of Paper I. 
Henceforth, $\phi_\ru$\ will be referred to as "non-common path aberration" (NCPA), as it exists in the WFS beam path, but not the SC beam path.
The Jones pupil matrix $A^\rk_w$\  accounts for known polarization effects, and it is assumed that unknown polarization effects between the BS and the WFS are negligible.
Of course, Eq.~(\ref{Upsilon_wB}) could be generalized to include more sophisticated effects, such as unknown polarization (leading to more unknowns) or Fresnel propagation, but it is likely that the most important character of the operator is captured here.
Assuming the validity of Eq.~(\ref{Upsilon_wB}), its inverse is given by:
\begin{equation}
\Upsilon^{-1}_{w,B}(\br_B,\br_w) = \delta\big( \br_B - \br_w \big) 
\big[ \mathbb{I} + A_w^\rk(\br_w)  \big]^{-1} \exp [  j \phi_\ru(\br_w)  ] \, .
\label{Upsilon_wB_inv}
\end{equation}
Combining Eqs.~(\ref{u_C-u_w-2}) and (\ref{Upsilon_wB_inv}) and performing the integration over $\br_w$ yields:
\begin{equation}
 u_C(\br_C,t) = \Upsilon_{C,B}(\br_C,\br_B)\big[ \mathbb{I} + A_w^\rk(\br_B)  \big]^{-1} 
  \exp[j \phi_\ru(\br_B)] u_w(\br_B,t) \, .
\label{u_C-u_w-1}
\end{equation}

Now, according to Eq.~(5) in Paper I, the wavefront sensor field $u_w(\br_w,t)$ admits the factorization:
\begin{equation}
u_w(\br,t) = \sqrt{I_\star \,} \exp j [ \beta k \balpha_\star \cdot \br +  \phi_\rw(\br,t) ] \breve{u}_w(\br,t) \, ,
\label{u_w-factor}
\end{equation}
where $\phi_\rw$\ is complex valued, and the $2 \times 1$\ polarization state vector $\breve{u}_w(\br,t)$ is normalized so that $\breve{u}^\rH_w(\br,t) \breve{u}_w(\br,t) = 1$, where the $^\rH$\ superscript indicates Hermitian conjugation.
For lack of a better term, $\phi_\rw$\ also will be called the "residual phase," despite the discussions in Sec.~\ref{OpticalSystem}.\ref{DM}.
Indeed, assuming the model in Eq.~(\ref{Upsilon_wB}), the true residual phase, $\phi_\rr$, and $\phi_\rw$\ are identical, and this nomenclature will be useful.
In Eq.~(\ref{u_w-factor}) the presence of the $\beta k \balpha_\star \cdot \br_w$ phase term, corresponding to the position of the star relative to the telescope pointing direction is part of the definition of $\phi_\rw$, and is included in this explicit form because SC will likely be more sensitive to the star position than the WFS due to the presence of the coronagraph. 
Inserting Eq.~(\ref{u_w-factor}) into Eq.~(\ref{u_C-u_w-1}) gives:
\begin{equation}
 u_C(\br_C,t) = \sqrt{I_\star \,} \Upsilon_{C,B}(\br_C,\br_B) \big[ \mathbb{I} + A_w^\rk(\br_B)  \big]^{-1}   \breve{u}_w(\br_B,t) 
  \exp j[ \phi_\ru(\br_B) +  \beta k \balpha_\star \cdot \br_B + \phi_\rw(\br_B,t) ]  \, ,
\label{u_C-u_w}
\end{equation}
which is fundamental to these developments.

An unavoidable complication is the fact that the signal measured by the SC is a function of the polarization state of the beam incident on the WFS, $\breve{u}_w(\br_w,t)$, which must somehow be determined.
In conventional AO systems, {\it la raison d'\^etre} of the WFS is to measure $\phi_\rw(\br_w,t)$, and, in principle, these measurements can be back-propagated to obtain an estimate of the atmospheric modulation $\phi_\ra(\br_0,t)$, from which, in turn,  $\breve{u}_w(\br_w,t)$\ can be estimated from Eq.~(\ref{u_w_complete}).
The back propagation requires ignoring the (presumably) small aberration $\phi_\ru$\ and other unknown aberrations in the optical system, which may well be adequate for this purpose.
An alternative, or complementary, approach to back-propagation would be to use a WFS that measures the polarization state, as suggested by \cite{Breckinridge15} in context of using such measurements to adaptively correct for the effects of polarization aberration.  
Estimates of $\breve{u}_w$\ based on back-propagation of $\phi_\rw$, may have signal-to-noise advantages over direct measurement, as the time-variable polarization effects are likely to be small.
It is important to note that, if the WFS employs a noiseless detector, measurements of higher spatial and temporal bandwidth should not compromise the performance of the AO system because the data can be binned in space and time for the purposes of running the AO servo loop.
To the best of this author's knowledge, all current WFSs measure the real part of the phase of the wavefront, but WFS concepts that measure the phase, amplitude and polarization state should be investigated.

\subsection{Aberrations Downstream of the BS}\label{downstream}

Eq.~(\ref{u_C-u_w}) accounts for aberrations upstream of the BS by virtue of being in terms of the field impinging on the WFS, $u_w$\ and including an NCPA term $\phi_\ru$.
It also accounts for for additional aberrations that are downstream of the BS on the way to the SC, but this is implicit.  
Below, we make this dependence explicit.
Recall that Eq.~(\ref{u_C-u_w-1}) expresses the SC field $u_C$\ in terms of WFS field $u_w$\ and $\Upsilon_{C,B}$, which propagates the field from the BS to the SC.
Aberrations downstream of the BS can be taken into account using the formalism in Sec. 4.A of Paper I, but instead of needing to consider all of the surfaces from $0$ to $C$ (the SC detector), we only need to consider those between the BS (index $B$) and the SC.
Similarly to Eq.~(39) of Paper I, the operator $\Upsilon_{C,B}$\ can be split into known and unknown parts as:
\begin{multline}
\Upsilon_{C,B}(\br_C,\br_B) \approx 
\Upsilon^\rk_{C,B} \big( \br_C,\br_B \big) \; + \\
\sum_{k=B}^{C-1} \left( \prod_{l=k+1}^{C-1}  \Upsilon^\rk_{l+1,l} \big( \br_{l+1},\br_{l} \big)      \right) 
 \Upsilon^\rk_{k+1,k} \big (\br_{k+1},\br_{k} \big)  \tilde{A}^\ru_{k}(\br_k) 
 \left( \prod_{l=0}^{k-1}  \Upsilon^\rk_{l+1,l}\big( \br_{l+1},\br_{l} \big) \right) \, ,
\label{AberOperator}
\end{multline}
where the $\{ \tilde{A}^\ru_k (\br_k) \}$ is the unknown $2 \times 2$ matrix of aberration functions corresponding to optical surface $k$, and $\{ \Upsilon^\rk_{l,l+1} \}$\ are known operators.
As per Sec.~3C of Paper I,  $\tilde{A}^\ru_k (\br_k) $\ is easily specialized to the case of a complex-valued scalar aberration, $\phi_k^\ru(\br_k)$, as: $\tilde{A}_k^\ru = \mathbb{I}\big( j\phi_k^\ru(\br_k) - {\phi_k^\ru}^2 (\br_k)+\cdots  \big)$.
Eq.~(\ref{AberOperator}) is not an equality since it does not allow for the aberrant fields to interact with other aberrations, reminiscent of the Born approximation in scattering theory.
Multiple reflections are ignored as well.
The reader may notice that the propagation operators in Eq.~(39) of Paper I contain a $\bm(t)$\ argument, allowing dependence on the DM command positions, but this is not needed here because BS is downstream of the DM.
It is likely permissible to omit the $k=B$\ term since it corresponds to the BS, which is conjugate to the WFS, so this aberration would be (nearly) indistinguishable from the non-common path error $\phi_\ru$, which could be re-defined to include it.

\section{The WFS Measurements}\label{WFS}

Sec.~\ref{OpticalSystem}.\ref{WFSandSC} contains developments that allow the polarimetric image in the SC to be expressed in terms of field that is incident on the WFS.  This section treats WFS measurements themselves, which later will allow a rigorous formulation of the statistical inference problem.
As was discussed, the field impinging on the WFS, $u_w(\br)$, can be factored as per Eq.~(\ref{u_w-factor}), in terms of the polarization state $\breve{u}_w(\br,t)$\ and the complex-valued phase $\phi_\rw(\br,t)$.  
The other phase factor in Eq.~(\ref{u_w-factor}), $\beta k \balpha_\star \cdot \br$, accounts for the pointing error, $\balpha_\star$.
The value of $\balpha_\star$\ can be defined to be the value that satisfies $ \big\langle \beta k \balpha_\star \cdot \br +  \Re [ \phi_\rw(\br,t)] \big\rangle_{\tau_\mathrm{G}} = 0$, where $\Re[ \;]$\ indicates the real part, and $  \langle \:  \rangle_{\tau_\mathrm{G}}$ denotes a time average over many Greenwood times $\tau_\rG$, but not so many that the total period approaches the timescale of dynamical evolution of the optical system $\tau_\rD$, as per the discussion in Sec.~2A of Paper I.

In order to contain the scope of this discussion, it will be assumed that polarization state $\breve{u}_w(\br,t)$\ is known, although this assumption must be evaluated with care.  
Standard statistical techniques take its uncertainty into account, but this would require knowing error covariance between its measurement and the measurement of $\phi_\rw$, which would depend greatly on the hardware and the technique for determining $\breve{u}_w(\br,t)$.
Thus, this issue will need to be revisited at a later time.

\subsection{WFS Bias, Gain and Noise}

The estimate of the "residual phase" $\phi_\rw(\br,t)$, $\hat{\phi_\rw}(\br,t)$, will be a function of the WFS measurements, $\bw(t_i) = \{w_k(t_i)\}, \; 0 \leq k < W$, where the WFS outputs a vector of $W$\ values at time $t_i$, in which $t_i$\ is a discrete variable corresponding to precisely recorded time-stamp, 
Note that $\bw$\ exists only on the discrete set of time indices $\{ t_0,\dots,t_{T-1}\}$.
The ideal value of $\bw(t_i)$, $\bw^\mathrm{ideal}(t_i)$, is the vector measurements that would made by a perfect WFS with no noise, biases or calibration errors.
The ideal coefficients are related to the measured ones $\bw(t_i)$ via:
\begin{align}
\bw^\mathrm{ideal}(t_i) & = \bw(t_i) + \bw^\mathrm{noise}(t_i) + \bw^\mathrm{bias}+ 
\bG \bw^\mathrm{ideal}(t_i)  \label{ideal_measurement-1} \\
& \approx \bw(t_i) + \bw^\mathrm{noise}(t_i) + \bw^\mathrm{bias}+ 
\bG \bw(t_i)
\label{ideal_measurement}
\end{align}
where $\bG$ is the $W\times W$\ gain matrix (with elements $\{ g_{ik} \}$), containing the uncalibrated part of the gain, $\bw^\mathrm{bias}$\ is the uncalibrated part of the bias, and $ \bw^\mathrm{noise}(t_i) $\ is a zero-mean stochastic process representing noise in the WFS measurement. 
Sources of noise may include photon-counting statistics and readout noise in the WFS detector.
It is assumed that best known calibration of the gain and bias are already incorporated into the measurement $\bw$, so that $\bG$\ and  $\bw^\mathrm{bias}$\ only represent presumably small corrections that need to be determined during the science observation.
It is important to note that $\bG$\ and  $\bw^\mathrm{bias}$\ are assumed to vary on the dynamical time-scale $\tau_\rd$ (as per Sec.~2A of Paper I).
Eq.~(\ref{ideal_measurement-1}) also presumes that the uncalibrated part of the gain is linear.
Eq.~(\ref{ideal_measurement-1}) cannot be put into practical use because it contains the unknowable $\bw^\mathrm{ideal}$ on the right-hand side.  
There is little choice but to make the approximation $\bG \bw^\mathrm{ideal}(t_i) \approx \bG \bw(t_i)$, the consequences of which diminish as the calibration improves and the unknown part of the gain, $\bG$, decreases.  

The entire point of millisecond imaging is to avoid integrating the signal over timescales that effectively average over the turbulence in the atmosphere, so turbulent averages are not considered, except as they relate to finite temporal bandwidth of the WFS, and such effects are treated later.
Then, from this point of view, the only stochastic element of Eq.~(\ref{ideal_measurement}) is WFS noise, which is captured in $\bw^\mathrm{noise}$, and all statistical operations will be carried out with respect to that process.  
The expected value of $\bw$\ is given by: 
\begin{align}
\mathrm{E} \big[ \bw(t_i) \,  | \,  \phi_\rw \big] &= (\mathbb{I} + \bG)^{-1}[\bw^\mathrm{ideal}(t_i) - \bw^\mathrm{bias} ] 
\label{w_mean-1} \\
& \approx (\mathbb{I} - \bG) [\bw^\mathrm{ideal}(t_i) - \bw^\mathrm{bias} ] \label{w_mean-2} \\
& \approx  \bw^\mathrm{ideal}(t_i) - \bw^\mathrm{bias} , 
\label{w_mean}
\end{align}   
in which $\mathrm{E}[\; ]$ is the expectation operator, and the 
''$ | \;  \phi_\rw(\br_0,t) $" is included to emphasize that the turbulent average is not being taken.
The approximation used to obtain Eq.~(\ref{w_mean-2}) relies on the hope that the uncalibrated gain $\bG$\ is "small" enough so that $(\mathbb{I} + \bG)^{-1} \approx \mathbb{I} - \bG $ [\citenum{BernsteinMatrixMath}, Prop.~11.3.10].
Similarly, the covariance of $\bw(t_i)$ is given by:
\begin{align}
\mathrm{cov} & \big[ \bw(t_i),\bw(t_i) \, |  \, \phi_\rw \big]   \nonumber \\
& =  (\mathbb{I} + \bG)^{-1} 
\mathrm{E}\big[\bw^\mathrm{noise}(\bw^\mathrm{noise})^\rH \big]
(\mathbb{I} + \bG)^{-\mathrm{H}}
  \label{w_covar-1} \\
  & \approx  (\mathbb{I} - \bG) 
\mathrm{E}\big[\bw^\mathrm{noise}(\bw^\mathrm{noise})^\rH \big]
(\mathbb{I} - \bG^\mathrm{H})
  \label{w_covar-2} \\
  & \approx  \mathrm{E} \big[\bw^\mathrm{noise}(\bw^\mathrm{noise})^\rH \big]
 \: \equiv \: \bC_{\bw} \, ,
 \label{w_covar}
\end{align}
thus defining the WFS noise covariance matrix $\bC_\bw$.

An ideal measurement $\bw^\mathrm{ideal}(t_i)$\ would lead to an ideal reconstruction of the residual phase $\phi'_\rw(\br,t_i)$, which is a temporally and spatially filtered version of the true value $\phi_\rw(\br,t)$.
The relationship between $\bw^\mathrm{ideal}(t_i)$\ and $\phi'_\rw(\br,t_i)$,  can be stated symbolically as $\bw^\mathrm{ideal}(t_i) \models \phi'_\rw(\br,t_i)$, where the $\models$\ symbol can be read as the word ''models.''  
The estimate of the residual phase $\hat{\phi_\rw}(\br,t_i)$\ must be a function of the real-world WFS data $\bw(t_i)$, so we also have the relationship $\bw(t_i) \models \hat{\phi_\rw}(\br,t_i)$.
As $\hat{\phi_\rw}$ inherits all of the physical limitations of the WFS, it must also be the estimate of $\phi'_\rw$, so $\hat{\phi_\rw} = \hat{\phi'_\rw}$.
The difference $\big[ \phi'_\rw(\br,t_i) - \hat{\phi_\rw}(\br,t_i)\big]$\ is a wavefront function corresponding to $\big[\bw^\mathrm{ideal}(\br,t_i) - \bw(\br,t_i)\big]$, and one may write $\big[\bw^\mathrm{ideal}(\br,t_i) - \bw(\br,t_i)\big] \models \big[\phi'_\rw(\br,t_i) - \hat{\phi_\rw}(\br,t_i)\big]$.
Similarly, each vector in Eq.~(\ref{ideal_measurement}) models a corresponding wavefront function:
\begin{eqnarray}
\bw^\mathrm{ideal}(t_i) & \models & \phi'_\rw(\br,t_i) \label{w_ideal_phi_prime} \\
\bw(t_i) & \models & \hat{\phi_\rw}(\br,t_i) \label{w_model} \\
\bw^\mathrm{noise}(t_i) & \models & \phi_\rn(\br,t_i) \\
\bG \bw(t_i) & \models & \phi_\rg(\br,t_i) \, . \label{gain_model} \\
\bw^\mathrm{bias} & \models & \phi_\rb(\br) \label{bias_model} 
\end{eqnarray}
As is the case with $\bw$, $ \phi'_\rw, \; \hat{\phi_\rw},\;  \phi_\rn $\ and  $\phi_\rg $ exist only on the discrete set of time indices $\{ t_0,\dots,t_{T-1}\}$, while the bias wavefront $ \phi_\rb$\ is not a function of time.
The problem of mapping these vectors onto their respective phase functions is the subject of Sec.~\ref{WFS}.

\subsection{Mapping the WFS into Continuous Time}\label{Mapping}

The phase of the wavefront impinging on the WFS $\phi_\rw(\br,t)$\ exists in continuous time, but the functions in  Eqs.~(\ref{w_ideal_phi_prime}) through (\ref{gain_model}) exist only on the set of discrete time $\{ t_i \}$ due to the fact that they are tied to the discrete-time output of the WFS.  
In order to examine more closely the relationship between the $\phi_\rw(\br,t)$ and the WFS output, consider that Eq.~(\ref{ideal_measurement}) and Eqs.~(\ref{w_ideal_phi_prime}) through (\ref{bias_model}) imply:
\begin{equation}
\phi'_\rw(\br,t_i) = \hat{\phi_\rw}(\br,t_i) + \phi_\rn(\br,t_i) + \phi_\rb(\br) +  \phi_\rg (\br,t_i) \, .
\label{phi_prime}
\end{equation}
As $\phi_\rw$ is continuous in space and time, it contains components that the WFS does not measure.
Let these components be represented by the continuous time signal $ \phihf(\br,t) $, where the ''hf" subscript is intended to remind the reader of the high frequency content of this function.
A precise definition $\phihf$\ is given later in Eq.~(\ref{blind_condition}).
It stands to reason that $\phi_\rw$\ should be the sum of $\phihf$ and $\phi_\rw'$, but first the $+$\ operator must be overloaded to allow the addition of discrete time and continuous time functions.
Let $\varphi(\br,t)$\ be a complex function existing on $\mathcal{D} \times \mathbb{R} $, where $\mathcal{D}$\ is the unit disk and $\mathbb{R}$\ is the real number line, and let $\psi(\br,t_i)$\ be a complex function existing on $\mathcal{D} \times \mathbb{Z}$, where $\mathbb{Z}=\{0,\dots,T-1\}$, which is isomorphic to the set of time indices $\{ t_i \}$.
So, $\varphi:  \mathcal{D} \times \mathbb{R} \mapsto \mathbb{C}$\ and $\psi: \mathcal{D} \times \mathbb{Z} \mapsto \mathbb{C}$, where $\mathbb{C}$\ is the set of complex numbers. 
The "$+$" operator can be overloaded to include the definition: 
\begin{equation}
\varphi(\br,t) + \psi(\br,t_i)
\equiv \varphi(\br,t) + \sum_{i=0}^{T-1} \psi(\br,t_i)\, \mathrm{rect}\left( \frac{t_i - t}{t_i - t_{i-1} }  \right) \, ,
\label{+_def}
\end{equation}
resulting in function that exists in continuous time. 
In Eq.~(\ref{+_def}), the rectangle function $\mathrm{rect}(t) = 1 $\ if $0 \leq t < 1$\ and $ = 0$\ otherwise, and $t_{-1}$\ is the time that the first exposure begins.
Of course, this interpolation scheme also serves to define an overloaded "$-$" operator, too.
Eq.~(\ref{+_def}) uses causal square-wave interpolation, which causes high-frequency components to be present in the sum.   
In the future, alternative interpolation functions which avoid this situation may be considered.

The sum operation defined in Eq.~(\ref{+_def}) allows one to add $\phihf$\ to $\phi_\rw'$:
\begin{equation}
\phi_\rw(\br,t) = \phi_\rw'(\br,t_i) + \phihf(\br,t) \, ,
\label{phihf}
\end{equation}
which serves as an alternative definition of $\phihf$.
Note that $\phihf$\ has no specific relationship to the DM servo loop and its bandwidth characteristics, as it only relates to the bandwidth of the WFS measurements themselves.
Indeed, $\phihf$\ would likely be only minimally affected by whether or not the DM servo loop is operating (as the controller cannot respond to fluctuations that the WFS does not measure).
Adding $\phihf(\br,t)$\ to both sides of Eq.~(\ref{phi_prime}) then gives a continuous time expression for the residual phase:
\begin{equation}
\phi_\rw(\br,t) = 
 \hat{\phi_\rw}(\br,t_i) + \phi_\rn(\br,t_i) + \phihf(\br,t) + \phi_\rb(\br) + \phi_\rg(\br,t_i)  \, .
\label{residual_phase_full}
\end{equation}

\subsection{Chromatic Error}\label{ChromaticError}

If the WFS operates at a wavelength that is different from the SC, then one must consider various effects of atmospheric dispersion including chromatic anisoplanitism and path length error [\citenum{Devaney_ChromaticAO,Jolissaint_ChromaticAO}].   
Let us define the chromatic error $\phi_\rc(\br,t)$, so that the wavefront at the SC wavelength impinging on the WFS is given by $\phi_\rw(\br,t) + \phi_\rc(\br,t)$.
In general, the polarization state of the light impinging on the WFS will depend on the wavelength of consideration, as well, so that instead it is given by $T_\rc(\br,t) \breve{u}_w(\br,t)$, where $T_\rc(\br,t)$\ is a matrix that accounts for this effect.   
While the rest of the discussion in this article could carry $ \phi_\rc(\br,t)$\ and $T_\rc(\br,t)$, meaningful treatment of uncertainties they imply is beyond the scope of this already somewhat complicated discourse and will have have be deferred to later work.

\subsection{NCPA and WFS Bias and Gain}\label{NCPA}

Eq.(\ref{residual_phase_full}) may be substituted into Eq.~(\ref{u_C-u_w}) to find the star's field at at the location $\brho$\ on the SC detector in terms of the WFS measurements:
\begin{multline}
u_C(\brho,t) = \sqrt{I_\star \,} \,  \Upsilon_{C,B}(\brho,\br_B) \big[ \mathbb{I} + A_w^\rk(\br_B)  \big]^{-1}  \breve{u}_w(\br,t) \; \times \\ 
\exp  j \big[ k \beta \balpha_\star \cdot \br   + \hat{\phi_\rw}(\br,t_i)  
+ \phi_\rn(\br,t_i)  + \phihf(\br,t) 
+ \phi_\ru(\br) + \phi_\rb(\br) + \phi_\rg(\br,t_i) \big] \, ,
\label{u_star_full_resid_phase-1} 
\end{multline}
where $\brho = \br_C$\ is the 2D spatial coordinate in the detector plane.
Note that in Eq.~(\ref{u_star_full_resid_phase-1}) the NCPA, $\phi_\ru$, and the WFS bias wavefront, $\phi_\rb$, have exactly the same form and are summed together.
Therefore, in this formulation, the wavefront function corresponding to the WFS bias behaves exactly like NCPA and it is impossible to distinguish the two (except, perhaps, by assumptions involving timescales of variability).
Thus it makes sense to simply treat them as a single aberration function and redefine the NCPA to include the effect of the WFS bias:  $\phi_\ru(\br) + \phi_\rb(\br) \rightarrow \phi_\ru(\br)   $, symbol overloading not withstanding.  
Then, Eq.~(\ref{u_star_full_resid_phase-1}) becomes:
\begin{multline}
u_C(\brho,t) = \sqrt{I_\star \,} \,  \Upsilon_{C,B}(\brho,\br) \big[ \mathbb{I} + A_w^\rk(\br)  \big]^{-1}  \breve{u}_w(\br,t)  \; \times \\ 
\exp  j \big[ k \beta \balpha_\star \cdot \br   + \hat{\phi_\rw}(\br,t_i)  
+ \phi_\rn(\br,t_i)  + \phihf(\br,t) 
+ \phi_\ru(\br) +  \phi_\rg(\br,t_i) \big] 
\label{u_star_full_resid_phase}
\end{multline}

\section{The SC Image}\label{SC_Image}

Eq.~(\ref{AberOperator}) can be inserted into Eq.~(\ref{u_star_full_resid_phase}) to obtain an expression for the science camera field that also includes unknown aberrations downstream of the BS:
\begin{multline}
u_C(\brho,t) \approx \sqrt{I_\star \,} 
\Bigg\{ \: \Upsilon^\rk_{C,B}(\brho,\br) \exp  j \big[  \phi_\ru(\br) +  \phi_\rg(\br,t_i) \big] \; + \\  
\hspace{-5mm}
\sum_{k=B}^{C-1} \left( \prod_{l=k+1}^{C-1}  \Upsilon^\rk_{l+1,l} \big( \br_{l+1},\br_{l} \big)      \right) 
 \Upsilon^\rk_{k+1,k} \big (\br_{k+1},\br_{k} \big)  \tilde{A}^\ru_{k}(\br_k) 
\left( \prod_{l=0}^{k-1}  \Upsilon^\rk_{l+1,l}\big( \br_{l+1},\br_{l} \big) \right)
  \\ \Bigg\} 
 \exp  j \big[ k \beta \balpha_\star \cdot \br   + \hat{\phi_\rw}(\br,t_i)  
+ \phi_\rn(\br,t_i)  + \phihf(\br,t)  \big] 
 \big[ \mathbb{I} + A_w^\rk(\br)  \big]^{-1}  \breve{u}_w(\br,t) \, ,
\label{u_star_full}
\end{multline}
Formally, the "unknowns" in Eq.~(\ref{u_star_full}) are the functions $\phi_\ru$, $\phi_\rg$ and the $\{ \tilde{A}^\ru_{k}\}$.
Note that cross-terms involving products of the $\{ \tilde{A}^\ru_{k}\}$\ and $\exp( j \phi_\ru)$\ or $\exp( j\phi_\rg)$\ have been dropped, as they are second order.
The reader is also reminded that the discrete-time and continuous-time phase terms are summed according to the rule in Eq.~(\ref{+_def}).

The function $J_{\star C}(\brho,t) = u_C(\brho,t) \otimes u^*_C(\brho,t)$ is the polarimetric image of the star incident on SC detector.
Calculating this function using Eq.~(\ref{u_star_full}) results in a somewhat lengthy expression, even after dropping terms of 2\underline{nd} order and higher in $\phi_\ru$, $\phi_\rg$ and the $\{ \tilde{A}^\ru_{k}\}$:
\begin{multline}
J_{\star C}(\brho,t) \approx  
I_\star \, \Bigg\{ 
\Upsilon^\rk_{C,B} \big(\brho,\br_B \big) \otimes  \Upsilon_{C,B}^{\rk *} \big(\brho,\br'_B \big) 
\exp  j \big[  \phi_\ru(\br) - \phi^*_\ru(\br') +  \phi_\rg(\br,t_i)  +  \phi^*_\rg(\br',t_i) \big] \;
 +  \\
 \Upsilon_{C,B}^\rk \big(\brho,\br_B \big) \otimes 
\hspace{0mm} 
\Bigg[ \sum_{k=B}^{C-1} \left( \prod_{l=k+1}^{C-1}  \Upsilon^{\rk *}_{l+1,l} \big( \br'_{l+1},\br'_{l} \big)      \right) 
\Upsilon^{\rk *}_{k+1,k} \big (\br'_{k+1},\br'_{k} \big)  \tilde{A}^{\ru *}_{k}(\br'_k)  \left( \prod_{l=0}^{k-1}  \Upsilon^{\rk *}_{l+1,l}\big( \br'_{l+1},\br'_{l} \big) \right) \bigg] \\
+
\\ 
\hspace{0mm} \Bigg[   \sum_{k=B}^{C-1} \left( \prod_{l=k+1}^{C-1}  \Upsilon^\rk_{l+1,l} \big( \br_{l+1},\br_{l}\big)      \right)
\Upsilon^\rk_{k+1,k} \big (\br_{k+1},\br_{k} \big)   \tilde{A}^\ru_{k}(\br_k) \left( \prod_{l=0}^{k-1}  \Upsilon^\rk_{l+1,l}\big( \br_{l+1},\br_{l} \big) \right)  \Bigg] 
 \otimes  \Upsilon_{C,B}^{\rk *} \big(\brho,\br'_B \big) \Bigg\}
 \\
\times \big[ \mathbb{I} + A_w^\rk(\br)  \big]^{-1} \otimes  \big[ \mathbb{I} + A_w^{\rk *}(\br')  \big]^{-1} 
 \big[ \breve{u}_w(\br,t) \otimes \breve{u}^*_w(\br',t) \big] 
    \exp  \big[ j \Phi(\br,\br',t) \big] 
\label{OMG}
\end{multline}
in which 
\begin{equation}
\Phi(\br,\br',t) \;  \equiv  \;  k \beta \balpha_\star \cdot( \br - \br') 
 +  \hat{\phi_\rw}(\br,t_i) - \hat{\phi_\rw^*}(\br',t_i) \; + \\
 \phi_\rn(\br,t_i) - \phi_\rn^*(\br',t_i) + \phihf(\br,t) - \phihf^*(\br',t) \, ,
\label{bigPhi_def}
\end{equation}
and $\brho \equiv \br_C$.
Note that  Eq.~(\ref{OMG}) is linear in the functions $\{ \tilde{A}^\ru_k  \}$, $\phi_\rg$ and $\phi_\ru$, if the latter are small enough to linearize exponentials.
The integrals Eq.~(\ref{OMG}) over $\br$ and $\br'$\ are separable, greatly easing the burden of numerical computations, and that the $16$ integrals (each corresponding to an element of the $ \Upsilon(\brho,\br) \otimes \Upsilon^*(\brho,\br')$ matrix) may be calculated with $4$ numerical integrations.

The imaging problem that confronts us is extremely asymmetric in the sense that the starlight must be treated with all possible detail and care, while a much more simple model of the optical propagation is adequate to describe the planetary image.
Let $\Upsilon^\rp_{C,B}(\br_C,\br_B)$\ be the needed low-order optical model for the planetary light that propagates fields from the beam splitter to the SC detector, containing no unknown quantities.
Since the BS and the WFS are in conjugate pupil planes (assuming no magnification effects), the field in the BS plane is well-enough approximated by $u_B(\br,t) \approx u_w(\br,t)$.  
Then, using Eqs.~(11) and (23) from Paper I, and the fact that the $B$\ plane is conjugate to the $0$ (telescope entrance pupil) plane, one finds for the polarimetric planetary image: 
\begin{equation}
J_{\rp C}(\brho,t)  
  = \int \rd \balpha \,  S(\beta \balpha) \, \bigg\{
     \big[ \Upsilon^\rp_{C,B}(\brho,\br) \otimes \Upsilon^{\rp*}_{C,B}(\brho,\br') \big] 
    \exp  j \big[ k \beta \balpha \cdot( \br - \br') +  \phi_\rw(\br,t)   
    - \phi_\rw^*(\br',t)  \big] \bigg\}   \, ,
 \label{planet_intensity_scicam-1}
\end{equation}
where the polarimetric planetary image that we ultimately wish to estimate is given by $S(\balpha)$.
The scalar version of the quantity contained in the braces in Eq.~(\ref{planet_intensity_scicam-1}) was called the ''planetary intensity kernel" in [\citenum{Frazin13}].
Recalling that $\phi_\rb$\ has been absorbed into $\phi_\ru$, the expression for $\phi_\rw$\ from Eq.~(\ref{residual_phase_full}) may be substituted into Eq.~(\ref{planet_intensity_scicam-1}), resulting in:
\begin{multline}
J_{\rp C}(\brho,t)  
  = \int \rd \balpha \, S(\beta \balpha) \, \bigg\{
     \big[ \Upsilon^\rp_{C,B}(\brho,\br) \otimes \Upsilon^{\rp*}_{C,B}(\brho,\br') \big] \; \times 
     \\  
       \exp  j \big[ k \beta \balpha \cdot( \br - \br')  +  \hat{\phi_\rw}(\br,t_i)  
    - \hat{\phi_\rw^*}(\br',t_i) \; + 
     \phi_\rn(\br,t_i) - \phi^*_\rn(\br',t_i) + \phihf(\br,t) - \phihf^*(\br',t)     \big] \bigg\}  \, ,
 \label{planet_intensity_scicam}
\end{multline}
where the unknown WFS bias and gain terms, the unknown functions $\phi_\ru$\ and $\phi_\rg$\ have been dropped since they should be inconsequential for the planetary image.  
Additionally, the terms involving  $\phi_\ru$\ and $\phi_\rg$\ are of 2\underline{nd} order in unknown quantities to be determined via statistical inference.

Assume that the intensity measured by the SC can be expressed as some linear combination of the Stokes parameters, corresponding to a measurement operator $\bM$, and leading to a scalar-valued measured intensity, $ \bM \big(J_{\star C} + J_{\rp C}  \big)$. 
For example, if the camera measures the $2$\underline{nd} Stokes parameter only (diagonal polarization), then $\bM = [0, \, 0, \, 1, \, 0] \bQ$, where $\bQ$\ is the $4 \times 4$\ matrix that converts a coherency vector to a Stokes vector, as given in Sec.~2A of Paper I.
More commonly, the SC may be measuring something rather close to the total intensity, in which case $\bM \approx  [1, \, 0, \, 0, \, 0] \bQ$.
At a pixel with position $\brho_l$\ in the exposure with timestamp $t_i$, the measured intensity, $I_\rm$, is given by integrating the stellar and planetary coherency vectors over a period of $(t_i  - t_{i-1})$\ milliseconds:
\begin{equation}
I_\rm(\brho_l,t_i)  =   \nu(\brho_l,t_i) +  \int^{t_i}_{t_{i-1}} \mathrm{d} t \,  \bM\big[ J_{\star C}(\brho_l,t) + J_{\rp C}(\brho_l,t)  \big] \, ,
\label{measured_intensity}
\end{equation}
where $\nu(\brho_l,t_i)$ accounts for noise associated with the measurement of the intensity itself.
Note that in Eq.~(\ref{measured_intensity}) the timestamps $t_{i-1}$\ and $t_i$\ need not correspond to the timestamps of the WFS measurements, but introducing a second set of timestamps would complicate the notation and discussion.
That said, detector readout noise may well provide motivation for taking some exposures in the SC with longer duration than the WFS exposure times [\citenum{Frazin14}].

Expected sources of error included in $\nu$\ are readout noise and photon-counting (shot) noise.
Eq.~(\ref{measured_intensity}) ignores integration over the detector pixel, tacitly assuming that $J_{\star C}$\ and $J_p$\ vary negligibly over the area of a detector pixel.
If the detector integration time is comparable to or shorter than the inverse Greenwood time, which is the AO correction timescale (see Paper I, Sec. 1), then the planetary image can be assumed to be constant over the time interval, and one can make the approximation: 
 \begin{equation}
 I_\rm(\brho_l,t_i)  \approx \nu(\brho_l,t_i) + \int^{t_i}_{t_{i-1}} \mathrm{d} t \, \bM J_{\star C}(\brho_l,t)  
+   (t_i - t_{i-1}) \bM J_{\rp C}(\brho_l,t_i)   \,  .
\label{measured_intensity_short} 
\end{equation}
Importantly, Eq.~(\ref{measured_intensity}) does not assume that the image of the star is constant over the interval, which will allow for treatment high-frequency variation of the stellar speckle.

\section{Semi-Analytical Reconstructor of the Residual Phase}\label{reconstruction}

In this section a reconstructor for the residual phase that should be useful for post-analysis will be discussed.  
The reconstruction will meet conditions of statistical optimality and will also prove to be helpful for expressing the covariance of $\phi_\rn(\br,t)$, as will be seen below.
The reconstruction method discussed here is not designed to minimize computation and is likely not suitable for implementation in closed-loop AO control.   

\subsection{Linear WFS Model}\label{WFS Model}

References [\citenum{Barrett07}] and [\citenum{Bechet09}] discuss the application of various estimation procedures to wavefront reconstruction from WFS data.
This section assumes an analytical model of the relationship between the vector of idealized WFS measurements, $\bw^\mathrm{ideal}(t_i)$, and the residual phase, $\phi_\rw(\br,t)$, in order to produce a semi-analytical expression for the reconstruction (or estimate) of $\phi_\rw(\br,t)$, denoted by $\hat{\phi_\rw}(\br,t_i)$. 
Assuming it is a linear device, the WFS measurement model may be expressed in terms of known linear functionals $\{ \chi_k \}$\ of the residual phase:
\begin{equation}
w^\mathrm{ideal}_k(t_i) = 
   \int \rd \br \,  \chi_k(\br) \int_{ t_{i-1}}^{t_i} \rd t \,  \phi_\rw(\br,t) \, ,
\label{WFS_model}
\end{equation}
where the corresponding function $\chi_k(\br)$\ maps the wavefront onto the complex numbers. 
For example, in the case of a Shack-Hartmann WFS, $ \chi_k(\br)  \approx \delta(\br - \br_k)  \nabla $, where $\br_k$\ is the location of the $k$\underline{th} lenslet, $\nabla$\ is the 2D gradient operator.  
Henceforth, the functions $ \{ \chi_k(\br) \}$\ will be referred to as the "modes" of the WFS.
Given an analytical form of the $\{ \chi_k(\br) \}$\ functions, such as in Eq.~(\ref{WFS_model}), the residual phase can be reconstructed with a semi-analytical pseudoinverse, optimally taking into account the noise statistics of the noise in the WFS, represented by $\bw^\mathrm{noise}$.

Let us first define an inner product on the space of WFS output values as $\langle \bw_1, \bw_2 \rangle \equiv \bw_2^\rH  \bw_1$.
Similarly, the inner product on the wavefront space is defined as:
 $
 \langle\phi_1(\br),\phi_2(\br)\rangle = \int \rd \br \,\phi^*_2(\br) \phi_1(\br)
 $. 
Now, let $\bchi$\ be a weighted operator mapping any function $\varphi(\br)$\ in wavefront space to WFS output space:
\begin{equation}
\bchi \varphi \equiv  \bC^{-1/2}_\bw
 \int   \rd \br \, 
\left[\begin{array}{l}
\chi_0(\br) \\
\vdots \\
\chi_{W-1}(\br)
\end{array} \right]
\varphi(\br)  \, ,
\label{bchi_def}
\end{equation}
where $\bC_{\bw} $\ was defined in Eq~(\ref{w_covar}), and $\bC^{-1}_\bw = \bC^{-\rH/2}_\bw \bC^{-1/2}_\bw$, in which $\bC^{-\rH/2}_\bw \equiv (\bC^{-1/2}_\bw)^\rH$.
In Eq.~(\ref{bchi_def}), the object inside the brackets $[ \, ]$\ is a $W \times 1$\ vector of continuous functions, also known as a "quasimatrix" \cite{Townsend_quasimatrix14}.
The reason for inclusion of $\bC_\bw^{-1/2}$\ as a pre-factor in definition of $\bchi$\ is because the weighting is required for statistical optimality as shown below in Eq.~(\ref{w_cost}).
With these definitions, Eq.~(\ref{WFS_model}) can be restated as:
\begin{equation}
\bw^\mathrm{ideal}(t_i) = \bC^{1/2}_\bw    \bchi   \int_{t_i}^{t_{i+1}} \rd t \,  \phi_\rw(\br,t) \, .
\label{w_phi'}
\end{equation}
The actual measurements, $\bw(t_i)$, and the estimate of the residual phase, $\hat{\phi_\rw}(\br,t_i)$, are related  by a rather similar equation:
\begin{equation}
\bw (t_i) = \bC^{1/2}_\bw    \bchi  \, \hat{ \phi_\rw}(\br,t_i) \, ,
\label{hat_w_phi}
\end{equation}
in which the time-integration is not needed since $\hat{ \phi_\rw}(\br,t_i)$\ is necessarily a discrete-time quantity, as per the discussion in Sec.~\ref{WFS}\ref{Mapping}.
Determining  $\hat{ \phi_\rw}(\br,t_i)$\ from $\bw(t_i)$\ is a matter of inverting Eq.~(\ref{hat_w_phi}), which can be achieved by first considering the generalized least-squares (GLS) cost function:
\begin{align}
\mathrm{cost}\big( \hat{\phi_\rw}(\br,t_i) \big) 
& \equiv 
  \big(\bw(t_i) -   \bC^{1/2}_\bw    \bchi  \, \hat{ \phi_\rw}(\br,t_i) \big)^\rH \bC_\bw^{-1}
\big(\bw(t_i) -   \bC^{1/2}_\bw    \bchi  \, \hat{ \phi_\rw}(\br,t_i) \big) \nonumber  \\
& =  
 \big( \bC^{-1/2}_\bw \bw(t_i) -      \bchi  \, \hat{ \phi_\rw}(\br,t_i) \big)^\rH 
\big( \bC^{-1/2}_\bw \bw(t_i) -    \bchi  \, \hat{ \phi_\rw}(\br,t_i) \big) \, .
\label{w_cost}
\end{align}

\subsection{Singular Value Decomposition}\label{SVD}

The cost function in Eq.~(\ref{w_cost}) has infinitely many minimizers due to the fact that $\hat{\phi_\rw}(\br,t_i)$\ is a continuous function of $\br$, and therefore infinite-dimensional, while it must be estimated from the $W$\ values contained in the WFS measurements $\bw$.
The classical solution to this common conundrum is to choose the least-squares solution of minimum norm (LSMN), as can be achieved with the singular value decomposition (SVD) \cite{Moon&Stirling}.   
The approach given here utilizes the SVD of the continuous-to-discrete operator, also called a "quasimatrix" \cite{Townsend_quasimatrix14}, $\bchi$, thus leveraging the analytical model functions $\{ \chi_k(\br) \}$ in Eq.~(\ref{WFS_model}).  
One advantage of this approach is that the covariance of the residual phase estimate has a convenient representation in terms of the $\{ \chi_k(\br) \}$, as will be seen below in Eq.~(\ref{hat_phi_w_cov}).

Given the inner product definitions defined above, it is straightforward to show that the adjoint of the operator $\bchi$\ is the following $1 \times W$\ quasimatrix times a weight matrix:
\begin{equation}
\bchi^\dagger = \big[ \chi^*_0(\br), \; \dots \; , \chi^*_{W-1}(\br) \big] \bC^{-\rH/2}_\bw \; .
\label{bchi_adjoint}
\end{equation}
Note that although the $\bchi$\ is continuous-to-discrete mapping, the operator $\bchi \bchi^\dagger$\ is a positive definite $W \times W$\ matrix (assuming that the modes $\{ \chi_k(\br) \}$ are non-redundant):
\begin{equation}
\bchi \bchi^\dagger = 
\bC^{-1/2}_\bw
  \Bigg\{
\int  \rd \br \,  
\left[\begin{array}{l}
\chi_0(\br) \\
\vdots \\
\chi_{W-1}(\br)
\end{array} \right] 
 \big[ \chi^*_0(\br), \; \dots \; , \chi^*_{W-1}(\br) \big] \Bigg\}
 \bC^{-\rH/2}_\bw 
 \, ,
\label{bchi_bchi_adjoint}
\end{equation}
in which numerical evaluation of the integrals $\int \rd \br \, \chi_k(\br) \chi^*_l(\br)$\ should present few problems.
The eigenvalues of $\bchi \bchi^\dagger$ are $[\sigma_0^2,\dots,\sigma_{W-1}^2]$, and the corresponding matrix of orthonormal eigenvectors is $\bV = [\bv_0, \, \dots, \, \bv_{W-1}]$. 
The positive numbers $\{ \sigma_l \}$\   are the singular values of $\bchi$, which can be expressed in terms of its singular value decomposition \cite{Townsend_quasimatrix14}:
\begin{equation}
\bchi = \sum_{l=0}^{W-1} \sigma_l \bv_l \zeta^*_l(\br) \, .
\label{bchi_svd}
\end{equation}
 The adjoint of $\bchi$\ can also be expressed in terms of the SVD:
 \begin{equation}
\bchi^\dagger = \sum_{l=0}^{W-1} \sigma_l \zeta_l(\br) \bv^\rH_l \, .
\label{bchi_adjoint_svd}
\end{equation}
In Eqs.~(\ref{bchi_svd}) and (\ref{bchi_adjoint_svd}) the functions $\{ \zeta_l(\br) \} $\ are orthonormal eigenfunctions of $\bchi^\dagger \bchi$, also with eigenvalues $\{ \sigma^2_l \}$.
$\bchi$\ has the pseudoinverse:
\begin{equation}
\bchi^\ddagger = \sum_{l=0}^{W-1} \frac{I}{\sigma_l} \zeta_l(\br) \bv^\rH_l  \, .
\label{bchi_pinv}
\end{equation}   
Once the eigenvectors and eigenvalues of $\bchi \bchi^\dagger$, and have been determined numerically, then the eigenfunctions of $\bchi^\dagger \bchi$\ can be found by the relation $\sigma_l \zeta_l(\br) = \bchi^\dagger \bv_l$.
Thus, using Eq.~(\ref{bchi_adjoint}), the eigenfunctions $\{ \zeta_l(\br) \} $\ can be expressed in terms of the WFS model functions $\{ \chi_l(\br) \}$:
\begin{equation}
\zeta_l(\br) = \frac{I}{\sigma_l} \bchi^\dagger \bv_l = 
 \frac{I}{\sigma_l} 
 \big[ \chi^*_0(\br), \; \dots \; , \chi^*_{W-1}(\br) \big] 
 \bC^{-\rH/2}_\bw  \bv_l \, .
 \label{zeta_functions}
\end{equation}

The LSMN solution of Eq.~(\ref{w_cost}) is given by:
\begin{equation}
\hat{\phi_\rw}(\br,t_i) = \bchi^\ddagger \bC^{-1/2}_\bw \bw(t_i)  \, .
  \label{LSMN_resid_phase}
\end{equation}
Applying Eqs.~(\ref{bchi_pinv}) and (\ref{zeta_functions}) to Eq.~(\ref{LSMN_resid_phase}), one obtains:
 \begin{align}
\hat{\phi_\rw}(\br,t_i)  =  
& \: \big[ \chi^*_0(\br), \; \dots \; , \chi^*_{W-1}(\br) \big]  
 \bC^{-\rH/2}_\bw  \left[ \sum_{l=0}^{W-1} \frac{I}{\sigma^2_l}
   \bv_l \bv_l^\rH  \right] \bC_\bw^{-1/2}  \bw(t_i)
    \label{LSMN_resid_phase-1} \\
 \equiv & \: \bPsi(\br) \bw(t_i)
 \label{LSMN_resid_phase1}
 \end{align}
thus defining the $1 \times W$\ vector of functions $\bPsi(\br)$.   
Note that while the orthonormality of the $\{ \bv_l \}$\ implies that $\sum_{l=0}^{W-1} \bv_l \bv^\rH_l = \mathbb{I}$, the $1/\sigma^2_l$\ factor in the sum in Eq.~(\ref{LSMN_resid_phase-1}) prevents further simplification (unless $\sigma_l = \sigma = \mathrm{constant}$).
As can be seen in Eq.~(\ref{LSMN_resid_phase-1}), the reconstructed wavefront is now expressed in terms of the continuous-valued WFS modes $\{ \chi_l(\br) \}$.

Once the estimated matrix of uncalibrated gains, $\hat{\bG}$, has been obtained, one would expect $\bw(t_i) + \delta \bw(t_i)$, where $\delta \bw(t_i) = \hat{\bG}\bw(t_i)$\ to lead to a better estimate of $\phi'(\br,t_i)$.
Therefore, the pseudoinverse reconstruction in Eq.~(\ref{LSMN_resid_phase}) should be applied to $\bG \bw$\ in order to allow estimation of $\bG$.
Using Eqs.~(\ref{LSMN_resid_phase}) through (\ref{LSMN_resid_phase1}), the gain wavefront can be expressed as: 
\begin{eqnarray}
\phi_\rg (\br , t_i) &=&  \bchi^\ddagger \bC^{-1/2}_\bw \bG \bw(t_i)  \nonumber \\
  & = &
 \bPsi(\br) \bG \bw(t_i) \, .
\label{gain_expansion}
\end{eqnarray}

Eq.~(\ref{w_ideal_phi_prime}) does not give an explicit definition of $\phi_\rw'(\br,t_i)$, which corresponds to the wavefront measured by an ideal WFS, but any satisfying  Eq.~(\ref{w_phi'}) is mathematically admissible.  
Making use of Eq.~(\ref{bchi_def}) he solution of minimum norm is the most convenient, corresponding to the definition:
\begin{equation}
\phi_\rw'(\br,t_i) \equiv \bigl( \bC_\bw^{1/2} \bchi   \bigr)^\ddagger \bw^\mathrm{ideal}(t_i) \, ,
\label{phi'_def}
\end{equation} 
where $\bigl( \bC_\bw^{1/2} \bchi   \bigr)^\ddagger$\ is the pseudoinverse of $\bigl( \bC_\bw^{1/2} \bchi   \bigr)$.

\subsection{Statistics of the Wavefront Reconstruction}

The statistics of the reconstruction $ \hat{\phi_\rw}(\br,t_i) $\ are inherited from those of $\bw$.
Applying the reconstruction formula in Eq.~(\ref{LSMN_resid_phase}) to both sides of Eq.~(\ref{w_mean}), one can see that:
\begin{equation}
\mathrm{E}\big[  \hat{\phi_\rw}(\br,t_i)  \; | \; \phi_\rw(\br,t)  \big] =
 \bchi^\ddagger \bC^{-1/2}_\bw \big[ \bw^\mathrm{ideal}(t_i) -  \bw^\mathrm{bias}(t_i)     \big] \, .
\label{mean_phi_hat}
\end{equation}
Applying similar reasoning to Eq.~(\ref{w_covar}), one obtains a convenient result for the covariance of $\hat{\phi_\rw}(\br,t_i)$:
 \begin{align}
 \mathrm{cov}\big[\hat{\phi_\rw}(\br,t_i), & \hat{\phi_\rw}(\br',t_i)   \; | \; \phi_\rw(\br,t)  \big]  \\ 
&=  
  \left[ \displaystyle{\sum_{l=0}^{W-1} \frac{I}{\sigma_l}} \zeta_l(\br) \bv^\rH_l \right]  
 \bC^{-1/2}_\bw  
 \mathrm{cov}\big[ \bw(t_i),\bw(t_i) \; | \; \phi_\rw(\br,t)  \big]
  \bC^{-\rH/2}_\bw 
 \left[ \displaystyle{\sum_{l=0}^{W-1} \frac{I}{\sigma_m}}\bv_m \zeta^*_m(\br')  \right] 
  \nonumber \\
 & = 
  \displaystyle{\sum_{l=0}^{W-1} \frac{I}{\sigma^2_l}} \zeta_l(\br)\zeta^*_l(\br') \, ,
  \label{hat_phi_w_cov}
 \end{align}
 which makes use of Eq.~(\ref{bchi_pinv}), the definition of $\bC_\bw$, and the orthonormality of the singular vectors $\{ \bv_l \}$.
 Thus, the spatial covariance of the estimated wavefront, caused by the WFS noise, has a convenient expression in terms of the WFS modes.

\section{Stochastic Processes}\label{StochasticProcesses}

The objective of this series of papers is to provide a framework for statistical inference of the functions need to describe the important unknown properties of the telescope system and, simultaneously, the planetary image.
Statistical inference methods will need to account for stochastic processes $\phi_\rn$\ and $\phihf$, which are present in the phase of the wavefront, and $\bnu$\ which is noise in the SC image.
The purpose of this section is to describe these processes insofar as is possible with detailing specific hardware configurations.

\subsection{Measurement Noise Processes}\label{measurement_noise}

The optical system under consideration captures images at both the WFS and the SC, leading to the processes $\bw^\mathrm{noise}$\ in Eq.~(\ref{ideal_measurement}), and $\bnu$\ in Eq.~(\ref{measured_intensity}).

In the simplest case, $\bnu$\ represents photon counting noise and detector readout noise, both of which of have the property that their values at pixel index $l$\ and time index $t$\ are independently distributed (ID) processes, meaning that the value $\nu(\brho_l,t_i)$\ is statistically independent of $\nu(\brho_{l'},t_{i'})$, unless $l = l'$\ and $i = i'$.
The detector readout noise is usually modeled as Gaussian (though see, e.g., more sophisticated treatments by Basden et al. [\citenum{Basden_PhotonCountingCCD03,Basden_JATIS15}]) while the photon counting noise is given by a Poisson distribution.   
The Poisson distribution is given by:
\begin{equation}
\mathcal{P}_\mathrm{P}(m | \overline{m}) = \frac{\overline{m}^m}{m!}e^{-\overline{m} }\, ,
\label{Poisson_dist}
\end{equation}
where $\overline{m}$\ is the expected number of photo-counts and $m$\ is the observed number of photo-counts.
As $\overline{m}$\ increases above small values (say, 30), $\mathcal{P}_\mathrm{P}$\ quickly approaches a Gaussian distribution with variance equal to $\overline{m}$.  
As a matter of practice, the variance is often taken to be the observed number of photocounts, which becomes problematic for small observed values of $m$.  
Given a probability model for the readout noise $\mathcal{P}_\mathrm{R}(m)$, it can be combined with Poisson distribution via the usual convolution rule for summing random variables:
\begin{equation}
\mathcal{P}_\nu(m|\overline{m}) = 
\sum_{m'=0}^\infty  \mathcal{P}_\mathrm{R}(m-m') \mathcal{P}_\mathrm{P}(m' | \overline{m}) \, . 
\label{P_nu}
\end{equation}
Assuming that $\mathcal{P}_\mathrm{R}(n)$\ is zero-mean, then the expectation of Eq.~(\ref{P_nu}) is $\overline{m}$.
If $\overline{m}$\ (or, indeed, the measured number of photo-counts $m$) is small, then not many terms in the sums of Eqs.~(\ref{Poisson_dist}) and (\ref{P_nu}) are needed.
On the other hand, the number of expected photo-counts may be large enough so that the Gaussian approximation to $\mathcal{P}_\mathrm{P}$\ applies.
In the ''large'' count regime, and if $\mathcal{P}_\mathrm{R}$\ is a zero-mean Gaussian with variance $\sigma^2_\mathrm{R}$, then Eq.~(\ref{P_nu}) takes the particularly simple Gaussian form:
\begin{equation}
\mathcal{P}_\nu(m|\overline{m}) = 
\frac{I}{\sqrt{2\pi (\overline{m} +\sigma^2_\mathrm{R} ) \,}}
\exp \left[ - \frac{ (m - \overline{m})^2 }{2 (\overline{m} + \sigma^2_\mathrm{R}  ) }   \right]
 \, . 
\label{P_nu_Gaussian}
\end{equation}
The expectation of Eq.~(\ref{P_nu_Gaussian}) is $\overline{m}$. 
The statistical independence of $\nu(\brho_l,t_i)$\ and $\nu(\brho_{l'},t_{i'})$, leads to diagonal covariance structure (irrespective of the validity of Gaussian approximation to $\mathcal{P}_\mathrm{P}$):
\begin{equation}
\mathrm{cov} \big[ \nu(\brho_l,t_i),\nu(\brho_{l'},t_{i'}) \big] = \big[ \overline{m}(\brho_l,t_i) + \sigma^2_\mathrm{R}  \big] \delta_{l,l'}\delta_{i,i'} \, ,
\label{nu_cov}
\end{equation}
where $\delta$\ is the Kronecker delta.

Recall from Sec.~\ref{WFS} that the $\phi_\rn(\br_0,t_i)$ component of the residual phase error is purely the result of noisy WFS output.
It is proportional to $\bw^\mathrm{noise}(t)$\ and therefore inherits its statistics.
$\bw^\mathrm{noise}(t_i)$\ is itself the result of the photo-count and readout statistics in the WFS camera system and arguments similar to those surrounding Eqs.(\ref{Poisson_dist}) through (\ref{nu_cov}) apply, so it is natural to assume that 
$\bw^\mathrm{noise}(t_i)$ is statistically independent of $\bw^\mathrm{noise}(t_{i'})$.
Then,
\begin{equation}
\mathrm{cov}\big[ w^\mathrm{noise}_l(t_i), w^\mathrm{noise}_{l'}(t_{i'})  \big] =  
 \delta_{i,i'} \bC_\bw  \, .
\label{WFS_cov}
\end{equation}
Therefore, as discussed in Sec.~(\ref{reconstruction}), the covariance of $\phi_\rn$\ can be expressed in terms of the eigenfunctions of the operator $\bchi^\dagger \bchi$, given in Eq.~(\ref{zeta_functions}):
\begin{align}
\mathrm{cov}\big[\phi_\rn(\br,t_i),\phi_\rn(\br',t_{i'})   \big] & 
 = \mathrm{cov}\big[\phi_\rw(\br,t_i),\phi_\rw(\br',t_{i'})   \big]  \nonumber \\
& = \delta_{i,i'} \displaystyle{\sum_{l=0}^{W-1} \frac{I}{\sigma^2_l}} \zeta_l(\br)\zeta^*_l(\br') \, .
 \label{phi_n_total_cov}
\end{align}

\subsection{Turbulent Stochastic Processes}\label{turbulent_processes}

 Recall from Eq.~(\ref{phihf}) that the process $\phihf$\ expresses the error in estimated wavefront $\hat{\phi_\rw}$\ due to the  finite spatial and temporal bandwidth of the WFS, {\it i.e.}, the components of the residual phase $\phi_\rw$ that it simply cannot ''see."
 Unlike $\phi_\rn$\ and $\bnu$, whose statistics are inherited from photo-detection, the statistics of $\phihf$\ are inherited from those of atmospheric turbulence and the AO system.
 This discussion in ths section assumes that scalar part of the phase aberration imposed on the wavefront upstream of the WFS, $\phi_\rT$ in Eq.~(\ref{residual_phase_def}) is completely within the spatiotemporal bandwidth of the WFS and contributes nothing to $\phihf$.
 This assumption should be re-examined within the context of specific telescope models.
 
The WFS is ''blind" to a fluctuation $\phi(\br,t)$\ that satisfies the following condition [as per Eq.~(\ref{WFS_model})]:
\begin{equation}
 \left|  \int \rd \br \, \chi_l(\br) \int_{t_i}^{t_{i+1}} \mathrm{d}t \,  \varphi(\br,t) \right| << \sqrt{ (\bC_\bw)_{l,l} \, }  \: , \: 
  \forall \, l \in \{0,\dots,W-1\} \, .
\label{blind_condition}
\end{equation}
In other words, if the inner product of the fluctuation with any of the $W$\ WFS modes, integrated over the exposure time, is less than the noise level for that mode, then it is unobservable.
This is equivalent to stating that $\varphi(\br,t)$\ is in the null space of the $\bC^{1/2}_\bw \bchi \int \rd t$ and $\bchi \int \rd t$ operators.
Any fluctuation characterized spatial and/or temporal frequency that is sufficiently high would would satisfy the unobservability condition in Eq.~(\ref{blind_condition}).
For example, temporal frequencies greater than Nyquist temporal frequency of the WFS, would be hard to measure, though one should aware of aliasing effects in which a high frequency could have significant response.
Similar arguments apply to the spatial frequency of the fluctuation.

According to Eq.~(\ref{phihf}), $\phihf(\br,t) = \phi_\rw(\br,t) - \phi'_\rw(\br,t_i)$,  and it follows from Eqs.~(\ref{w_phi'}) and (\ref{phi'_def}) that:
\begin{align}
 \phihf(\br,t) & =  \phi_\rw(\br,t) -  \left( \bC^{1/2}_\bw \bchi \right)^\ddagger \bC_\bw^{1/2} \bchi \int_{t_i}^{t_{i+1}} \rd t \, \phi_\rw(\br,t) 
 \label{phihf_filter} \\
 & =   \int \rd \br  \int \rd t' \, \delta(\br - \br')\delta(t - t') \phi_\rw(\br',t')   
 - \left( \bC_\bw \bchi \right)^\ddagger \bC_\bw^{1/2} \bchi  \int_{t_i}^{t_{i+1}} \rd t \,   \phi_\rw(\br,t') 
\nonumber  \\ 
 & \equiv   \boldeta  \,  \phi_\rw(\br',t') \, ,
 \label{boldeta_def}
\end{align}
thus defining the operator $\boldeta$, which is a high-pass filter, both spatially and temporally. 
Note that the difference between the continuous-time and discrete-time terms is governed by the sum rule in Eq.~(\ref{+_def}).
Since (presumably) $\mathrm{E}[ \phi_\rw(\br,t) ] = 0$, it follows from Eq.~(\ref{boldeta_def}) that 
\begin{equation} 
\mathrm{E}[ \phihf(\br,t) ] = 0 \, .
\label{E_phihf}
\end{equation}
Then the covariance of $\phihf$ is  given by:
\begin{equation}
\mathrm{cov}\big[ \phihf(\br,t),\phihf(\br',t') \big] 
 =  \: \boldeta \,
\mathrm{cov}\big[ \phi_\rw(\br'',t''),\phi_\rw(\br''',t''') \big] \boldeta^\dagger \, ,
\label{phihf_cov}
\end{equation}
where $ \boldeta^\dagger$\ is the adjoint of $ \boldeta$.
The covariance of the residual phase, $\mathrm{cov}\big[ \phi_\rw(\br,t),\phi_\rw(\br',t') \big]$, is itself the subject of substantial importance in the AO community, and various treatments can be found in the literature [\citenum{Ellerbroek97,Ellerbroek99,Vogel14}].
Eq.~(\ref{phihf_cov}) presents a number of practical difficulties in that models of $\mathrm{cov}\big[ \phi_\rw(\br,t),\phi_\rw(\br',t') \big]$\ must be based on models of layered atmospheric turbulence as in [\citenum{Ellerbroek97,Ellerbroek99}] or analysis of specialized observations as in [\citenum{Vogel14}].
Determining the temporal statistics of the wavefront from atmospheric turbulence models requires the Taylor, or "frozen flow," assumption.
Under the Taylor assumption, a temporally static, but spatially structured layer of atmosphere, travels across the line-of-sight.
In a series of observations, Poyneer et al. [\citenum{Poyneer09}] found that only a minority of the ''controllable phase power" can be attributed to frozen flow.
Therefore, it seems that a preference should be given to experimental determination of the covariance as in [\citenum{Vogel14}], if possible.

The importance $\phihf$\ for our purposes is only due to its presence in the function $\Phi$\ in Eqs.~(\ref{bigPhi_def}) and (\ref{OMG}), which itself is integrated over the exposure time of the SC in Eq.~(\ref{measured_intensity}).
Note that, to fist order, $\int_0^{\delta t} \exp[j \phihf(\br,t)]  \rd t = \delta t + j \int_{\delta t}  \phihf(\br,t)  \rd t $.
Thus, there is significant motivation to the study the statistical properties of the time-integrated version of $\phihf$\ as they are likely to be more simple than Eq.~(\ref{phihf_cov}).
Consider integrating $\phihf$ over the exposure time in the SC:
\begin{equation}
\phihf^\downarrow(\br,t_i) \equiv \int_{t_i}^{t_{i+1}} \rd t \, \phihf(\br,t) \, , 
\label{phi_down_def}
\end{equation}
thus defining the discrete-time quantity $\phihf^\downarrow(\br,t_i)$.
Similarly, one can define a time-integrated residual phase:
\begin{equation}
\phi_\rw^\downarrow(\br,t_i) \equiv \int_{t_i}^{t_{i+1}} \rd t \, \phi_\rw(\br,t) \, .
\label{phi_w_down_def}
\end{equation}
Then, following steps similar to the ones that lead to Eq.~(\ref{boldeta_def}), one has:
\begin{align}
 \phihf^\downarrow(\br,t_i) & =  \phi^\downarrow_\rw(\br,t_i) -  \left( \bC^{1/2}_\bw \bchi \right)^\ddagger \bC_\bw^{1/2} \bchi \, \phi^\downarrow_\rw(\br,t_i) 
 \label{phihf_down_filter} \\
 & =   \int \rd \br  \, \delta(\br - \br') \phi_\rw(\br',t_i)   - \: 
 \left( \bC_\bw \bchi \right)^\ddagger \bC_\bw^{1/2} \bchi  \, \phi^\downarrow_\rw(\br,t_i) 
\nonumber  \\ 
 & \equiv   \boldeta ' \,  \phi^\downarrow_\rw(\br',t_i) \, ,
 \label{boldeta'_def}
\end{align}
thus defining the operator $\boldeta'$.
The difference between the two versions of the operator $\boldeta$\ in Eqs.~(\ref{boldeta_def}) and (\ref{boldeta'_def}) is that the latter does not involve integration over an exposure time.
Stated another way, the operator $\boldeta$\ is a high-pass filter both spatially and temporally, while $\boldeta'$\ is only a a high-pass spatial filter since it is already operating on a temporally filtered and discretized signal.
As was the case with Eq.~(\ref{E_phihf}), one has:
\begin{equation} 
\mathrm{E}[ \phihf^\downarrow(\br,t) ] = 0 \, .
\label{E_phihf_down}
\end{equation}
Similarly to Eq.~(\ref{phihf_cov}), one has:
\begin{equation}
\mathrm{cov}\big[ \phihf^\downarrow(\br,t_i),\phihf^\downarrow(\br',t_{i'}) \big] 
 =  \: \boldeta ' 
\mathrm{cov}\big[ \phi^\downarrow_\rw(\br'',t_i),\phi^\downarrow_\rw(\br''',t_{i'}) \big] {\boldeta '}^{ \dagger} \, .
\label{phihf_down_cov}
\end{equation}
Although Eq.~(\ref{phihf_down_cov}) looks rather similar to Eq.~(\ref{phihf_cov}), the former should indeed be simpler.  
Indeed, while $\mathrm{cov}\big[ \phihf(\br,t),\phihf(\br',t') \big] $\ is defined for any two times $t$ and $t'$, $\mathrm{cov}\big[ \phihf^\downarrow(\br,t_i),\phihf^\downarrow(\br',t_{i'}) \big] $, is only defined for discrete times $t_i$\ and $t_{i'}$.
Eq.~(\ref{phihf_down_cov}) only accounts for the properties of the time-integrated residual phase $\phi^\downarrow_\rw$, and has diminished high-frequency content, both spatially and temporally.
The spatial consequences of the temporal integration of $\phi_\rw$\ can be visualized by considering Taylor's frozen flow hypothesis, in which fluctuations with high spatial frequency also have high temporal frequency.

\section{Conclusions and Statistical Inference Strategy}

The polarimetric image intensity measured by the SC in Eq.~(\ref{measured_intensity}), is linear in the unknown functions that represent the NCPA, the uncalibrated WFS gains, the aberrations between the beam splitter and the SC, and the planetary image, denoted by $\phi_\ru(\br)$, $\bG$, $\{ \tilde{A}^\ru_k (\br_k) \}$, and $S(\balpha)$, respectively.\footnote{This requires linearizing the exponential containing $\phi_\ru$\ and $\phi_\rg$.}
Importantly, the SC image has been expressed in terms of the estimated wavefront $\hat{\phi_\rw}(\br,t_i)$, which itself is the result of a statistical optimal reconstruction based on a singular value decomposition of the analytical WFS observation modes.
This direct connection of the measured intensity in the SC to the reconstructed wavefront should
allow rigorous statistical inference procedures to utilize the WFS data stream.
The statistical inference problem must take into account several stochastic processes.
The first is $ \nu(\brho_l,t_i)$, which represents noise in the polarimetric image measured by the SC, and whose properties are relatively simple and can be treated via standard procedures for additive noise. 
The other stochastic processes are in the expression for the wavefront and provide more of a challenge to algorithm development.
Noise in the WFS manifests itself as the function $\phi_\rn(\br,t_i)$, whose covariance can be conveniently expressed in terms of the modes WFS, as shown by the singular value decomposition analysis in Sec.~\ref{reconstruction}.
Components of the wavefront that the WFS cannot detect due to their high temporal and spatial frequency, represented by $\phihf(\br,t)$, are included in the formulation, as well, and their statistics must be specified in future efforts.

This discussion has mostly avoided the complicated subject of chromatic error, which happens when WFS senses at a wavelength that is difference from that at which the SC operates.
The primary motivations for such schemes are benefits for operating WFS at shorter wavelengths (less noisy and expensive photodetection) and not spending science photons on wavefront sensing. 
However, chromatic error in AO takes on even more significance in the case of coronagraphy since it will reduce the efficiency with which the coronagraph extinguishes the starlight.
Thus, simulations are required to evaluate the benefits not operating the WFS and SC at a common wavelength.

It is worth remarking that it may be possible to recover some approximation of the phase error $\Delta \phi(\br,t) \equiv \phi_\rn(\br,t_i) + \phihf(\br,t)$ using the fact that most choices of $\Delta \phi(\br,t)$\ will lead to a predicted image that is inconsistent with the measured intensity $I_\rm$.
This problem bears considerable similarity to the phase retrieval problem discussed by Fienup [\citenum{Fienup93}].
The prospects of doing this successfully are complicated by the need to simultaneously estimate other unknown functions, however, the fact that they are expected to vary on much longer timescales may make it feasible.

Formally, the inference problem can be placed in canonical form for linear
linear statistical inference, i.e., $\by = \bH \bx + \bnu$, where
$\by$\ is the vector of measurements, $\bH$ is the system matrix,
$\bx$\ is the vector of unknowns to be estimated, and $\bnu = \{ \nu(\brho_l, t_i) \}$\
is the additive noise.  
However, it is important to emphasize that this is not the classical
least-squares problem for two reasons.   
The first is that the dependence of the elements in $\bH$\ on the
known but random WFS measument $\hat{\phi_\rw}(\br,t_i)$\  gives
the system a stochastic character (this is sometimes called the
``stochastic explanatory variable'' problem), in that the estimates will have
some dependence on the particular values of $\{ \hat{\phi_\rw}(\br,t_i) \}$\ that arise during the observation.
As there are to be a large number of millisecond exposures, this effect is likely negligible.
Much more important is the fact that the elements of $\bH$ are
functions of the stochastic functions $\phi_\rn(\br,t_i)$\ and
$\phihf(\br,t)$, effectively making $\bH$ ``noisy,'' which is
sometimes known as the ``errors in variables'' problem.   
One consequence of the errors in variables is that the standard
least-squares estimator is biased [\citenum{Kmenta_Econometrics}].
One framework for dealing with such problems that should be considered a series of generalizations of total least squares, as covered in the series of books edited by van Huffel and Lemmerling [\citenum{vanHuffel_TLS3}].

It is likely that a successful statistical inference solution will
exploit multi-scale (e.g., wavelet) representations of the unknown
functions, as well as other convenient structure, such as
quasi-separability that allows one to make a high-quality approximate
solution for the aberrations while taking the planetary image to be
zero, and then updating the aberration solution while simultaneously
estimating the planetary image.  
In order for the statistical inference method to be of practical use
in processing the large quantity of millisecond images, it must be
placed into sequential estimation framework, such as Kalman filtering.


\section*{Acknowledgments} 
The author thanks Olivier Guyon, Jim Breckinridge and Wes Traub for
enlightening discussions that improved this paper.  

\bibliography{exop.bib}   
\bibliographystyle{osajnl}   

\end{document}